\documentclass[groupedaddress,doublelines,10pt,twocolumn,final]{revtex4}%
\usepackage{amssymb}
\usepackage{amsmath}
\usepackage{amsfonts}
\usepackage{graphicx}%
\providecommand{\U}[1]{\protect\rule{.1in}{.1in}}

\begin{document}
\title{The Information-Complete Quantum Theory}
\author{Zeng-Bing Chen}
\email{zbchen@nju.edu.cn}
\affiliation{National Laboratory of Solid State Microstructures and School of Physics,
Nanjing University, Nanjing 210093, China}

\begin{abstract}
Quantum mechanics is a cornerstone of our current understanding of nature and
extremely successful in describing physics covering a huge range of scales.
However, its interpretation remains controversial since the early days of
quantum mechanics. What does a quantum state really mean? Is there any way out
of the so-called quantum measurement problem? Here we present an
information-complete quantum theory (ICQT) and the trinary property of
nature to beat the above problems. We assume that a quantum system's state
provides an information-complete description of the system in the trinary
picture. We give a consistent formalism of quantum theory that makes the
information-completeness explicitly and argue that conventional quantum
mechanics is an approximation of the ICQT. We then show how our ICQT provides
a coherent picture and fresh angle of some existing problems in physics. The
computational content of our theory is uncovered by defining an
information-complete quantum computer.\newline

\noindent \textbf{Keywords}: Information-complete quantum theory, trinary
description, dual entanglement, dual dynamics

\end{abstract}
\maketitle



\section{Introduction}

The unease of understanding quantum theory (QT) began at the very beginning of
its establishment. The famous Bohr-Einstein debate \cite{EPR,Bohr} inspired a
lively controversy on quantum foundations. QT is surely an empirically
successful theory, with huge applications ranging from subatomic world to
cosmology. However, why does it attract such a heated debate over its whole
history? The controversial issues on quantum foundations mainly focus on two
aspects: ($Q1$) What does a wave function (or a quantum state) really mean?
($Q2$) Is the so-called quantum measurement problem
\cite{WZ-book,Omnes,Zurek,Sun,Sun-rev,PhyRep} really a problem? The first
axiom of the standard QT states that a system's wave function provides a
complete description of the system. But accepting the wave function as QT's
central entity, what is the physical meaning of the wave function itself? In
this regard, there are two alternatives that the quantum state might be either
a state about an experimenter's knowledge or information about some aspect of
reality (an \textquotedblleft epistemic\textquotedblright\ viewpoint), or a
state of physical reality (an \textquotedblleft ontic\textquotedblright%
\ viewpoint). A recent result \cite{PBR} on this issue seems to support the
reality of quantum states, yet with ongoing controversy
\cite{PBR-comment,Aaronson}.

On the other hand, the quantum measurement problem is perhaps the most
controversial one on quantum foundations. According to the orthodox
interpretation (namely, the Copenhagen interpretation \cite{Omnes}) of QT, a
quantum system in a superposition of different states evolves
deterministically according to the Schr\"{o}dinger equation, but actual
measurements always collapse, in a truly random way, the system into a
definite state, with a probability determined by the probability amplitude
according to the Born rule. When, where, and how the quantum state really
collapses are out of the reach of QT as it is either \textquotedblleft
uninteresting or unscientific to discuss reality before
measurement\textquotedblright\ \cite{Aaronson}.

Our classical world view implies that there exists a world that is objective
and independent of any observations. By sharp contrast, what is observed on a
quantum system is dependent upon the choice of experimental arrangements;
mutually exclusive (or complementary) properties cannot be measured accurately
at the same time, a fact known as the complementarity principle. In
particular, which type of measurements one would like to choose is totally a
\textit{free will} \cite{free-will} or a \textit{freedom of choice}
\cite{free-choi74,Bell-book,free-choiEX}. Such a freedom of choice underlies
the Pusey-Barrett-Rudolph theorem \cite{PBR} and the derivation of Bell's
inequalities \cite{free-choi74,Bell-book,free-choiEX,Bell}. However, one could
ask: What does a free will or a freedom of choice mean physically and particularly,
what is the physical system that encodes information about the free
will or freedom of choice?

Thus, in the orthodox interpretation classical concepts are necessary for the
description of measurements (which type of measurements to choose and the
particular measurement results for chosen measurement) in QT, although the
measurement apparatus can indeed be described quantum mechanically, as done by
von Neumann \cite{vonN-book,Wigner}. Seen from its very structure, quantum
mechanics \textquotedblleft contains classical mechanics as a limiting case,
yet at the same time it requires this limiting case for its own
formulation\textquotedblright\ \cite{Landau}. In this sense current QT has a
classical-quantum \textit{hybrid} feature. At a cosmological scale, the
orthodox interpretation rules out the possibility of assigning a wave function
to the whole Universe, as no external observer could exist to measure the Universe.

Facing with the interpretational difficulties, various interpretations on QT
were proposed by many brilliant thoughts, such as the hidden-variable theory
\cite{Bell-book,Bohm} (initiated by the famous Einstein-Podolsky-Rosen paper
\cite{EPR} questioning the completeness of QT), many-worlds interpretation
\cite{Everett,Tipler}, the relational interpretation
\cite{Rovelli,Rovelli-book}, and the decoherence theory \cite{Zurek}, to
mention a few. Thus, ``questions concerning the foundations of quantum
mechanics have been picked over so thoroughly that little meat is left''
\cite{Aaronson}. The discovery of Bell's inequalities \cite{Bell} (as well
as the related quantum nonlocality, questioned from the many-worlds
interpretation \cite{Tipler}) and the emerging
field of quantum information \cite{QuInf-book} might be among a few
exceptions. The recent development of quantum information science sparks the
information-theoretical understanding of quantum formalism
\cite{Hardy,Fuchs,i-causality,Colbeck-Renner}.

Logically, if one is looking for a quantum formalism that is universally valid,
one should not assume anything beyond the description of the formalism. Current QT does
not meet this internal logic requirement as the change of its central entity,
quantum states, is separated into two broken pieces---The transition from a probability amplitude
(related to an unitary evolution) to a probability (related to a non-unitary
measurement) has to require a macroscopic, thus classical, apparatus.
Inspired by the classical-quantum hybrid feature of current QT and the
above-mentioned interpretational progresses, here we present an
information-complete quantum theory (ICQT) by removing any classical
systems/concepts in our description of nature. The ICQT is based on the
\textit{information-completeness principle}: Any information must be carried and acquired 
by certain quantum systems that encode their complete information. In this 
way, quantum states represent the information-complete code of any
possible information that one might acquire. Here, what do we mean by ``information''
and how to acquire or measure information? In a genuinely quantum world (i.e., there is
no room for any classical system), measurement is simply interaction between two physical
systems \cite{Rovelli-book}. The interaction creates quantum entanglement between the
two systems; what entanglement encodes is \emph{the} information about the two systems---Entanglement between the two systems acquires information of them; no entanglement means no information.
Now it is ready to see that the information-complete physical system must be a composite system,
whose constituents interact to mutually acquire information; to be information-complete, the number
of the constituents has to be three. As argued below, the three constituents are an indivisible
``trinity''.

The aim of the present work is thus to suggest a new quantum structure that respects
information-completeness principle. After working out the information-completeness explicitly in our formalism, we show that information-complete physical systems, whose definition is
to be given below, are characterized by
a trinary description, dual entanglement pattern, the emergent dual Born rule, and dual
dynamics. As enforced by the information-completeness principle within
the trinary picture, an information-complete trinity is characterized by the indivisibility of its kinematics and dynamics; \emph{its complete information (i.e., the
physical predictions) is merely dual entanglement, which acquires a very universal role in our
information-complete quantum formalism}.
The computational content of our theory is uncovered by defining an
information-complete quantum computer with a trinary structure and potential of outperforming
conventional quantum computers. Moreover, we consider the possible conceptual
applications of our theory, hoping to shed new light on some existing problems
in physics. Seen from the ICQT, current QT, due to its classical-quantum hybrid feature, is \textit{not}
information-complete and thus suffers from interpretational difficulties.

\section{An information-complete description for finite-dimensional
systems}

The orthodox quantum measurement theory
\cite{WZ-book,Omnes,Zurek,Sun,Sun-rev,PhyRep} was proposed by von Neumann and
can be summarized as follows. For an unknown $d$-dimensional quantum state
$\left\vert \psi\right\rangle _{\mathcal{S}}$ of a quantum system
$\mathcal{S}$ to be measured, a measurement apparatus (\textquotedblleft a
pointer\textquotedblright) $\mathcal{A}$ is coupled to the system via a
unitary operator $\hat{U}_{\mathcal{SA}}(\hat{s},\hat{p})$. Here $\hat{s}$ is
system's observable whose eigenstate with respect to the eigenvalue $s_{j}$
reads $\left\vert j,\mathcal{S}\right\rangle $, namely, $\hat{s}\left\vert
j,\mathcal{S}\right\rangle =s_{j}\left\vert j,\mathcal{S}\right\rangle $
($j=1,2,...d$); $\hat{p}$ is the momentum operator which shifts pointer's
$\hat{q}$-reading ($[\hat{q},\hat{p}]=i$). Assuming that the pointer is
initialized in a \textquotedblleft ready\textquotedblright\ state $\left\vert
0,\mathcal{A}\right\rangle $ and expanding $\left\vert \psi,\mathcal{S}%
\right\rangle $ in terms of $\left\vert j,\mathcal{S}\right\rangle $ as
$\left\vert \psi,\mathcal{S}\right\rangle =\sum_{j}c_{j}\left\vert
j,\mathcal{S}\right\rangle $, then the system and the apparatus are mapped
into
\begin{equation}
\hat{U}_{\mathcal{SA}}(\hat{s},\hat{p})\left\vert \psi,\mathcal{S}%
\right\rangle \left\vert 0,\mathcal{A}\right\rangle =\sum_{j}c_{j}\left\vert
j,\mathcal{S}\right\rangle \left\vert q_{j},\mathcal{A}\right\rangle .
\label{preM}%
\end{equation}
To ideally measure $\hat{s}$, one has to assume that $\mathcal{A}$ must have
at least $d$ macroscopically distinguishable pointer\ positions (plus the
ready position corresponding to $\left\vert 0,\mathcal{A}\right\rangle $), and
the pointer state $\left\vert q_{j},\mathcal{A}\right\rangle $ and the
measured states $\left\vert j,\mathcal{S}\right\rangle $ have an one-to-one
correspondence. The above is the usual pre-measurement progress. The orthodox
interpretation of the measurement can only predict the collapse of a definite
state $\left\vert j,\mathcal{S}\right\rangle $ with a probability $\left\vert
c_{j}\right\vert ^{2}$ given by the probability amplitude $c_{j}$; the
collapse occurs in a truly random way. For latter convenience, we call
($\hat{s},\hat{p}$) as an observable pair. It is interesting to note that a
factorizable structure of the \textquotedblleft measurement
operation\textquotedblright\ $\hat{U}_{\mathcal{SA}}(\hat{s},\hat{p})$ was
discovered in the context of the dynamical approach to the quantum measurement
problem \cite{Sun,Sun-rev}.

To avoid the quantum measurement problem, here we take a key step by assuming
explicitly information-completeness, whose meaning will be clear below, in
our formalism of describing nature. For measuring information on $\mathcal{S}%
$, one could of course choose various bases, namely, entangle $\mathcal{S}$
and $\mathcal{A}$ in different bases. In current QT, the observer has the freedom
for this kind of choices. However, \textit{in the ICQT, we require that all
information, including the basis information, must be encoded by certain quantum
system to avoid any classical terms or concepts}.\ To this end, starting from a
separable state $\left\vert \psi,\mathcal{S}\right\rangle \left\vert
\phi,\mathcal{A}\right\rangle $, we introduce the third system, called the
\textquotedblleft programming system\textquotedblright\ ($\mathcal{P}$)
hereafter to encode the basis information of $\mathcal{S}$ and $\mathcal{A}$.
We assume that $\mathcal{P}$ has $D_{\mathcal{P}}$ dimensions spanned by
$D_{\mathcal{P}}$ orthogonal states, called programming states $\left\vert
r,\mathcal{P}\right\rangle $ ($r=0,1,...D_{\mathcal{P}}-1$), where
$D_{\mathcal{P}}$ is to be determined by information-completeness. Let us
define a unitary programming operation
\begin{equation}
\hat{U}_{\mathcal{P(SA)}}=\sum_{r=0}^{D_{\mathcal{P}}-1}\left\vert
r,\mathcal{P}\right\rangle \left\langle r,\mathcal{P}\right\vert \hat
{U}_{\mathcal{SA}}(\hat{s}_{r},\hat{p}_{r}), \label{progU}%
\end{equation}
which means that if $\mathcal{P}$ is in $\left\vert r,\mathcal{P}\right\rangle
$, then do an unitary measurement operation $\hat{U}_{\mathcal{SA}}(\hat{s}%
_{r},\hat{p}_{r})$ on $\mathcal{SA}$. Now suppose that $\mathcal{P}$ is
prepared in an initial state $\left\vert \chi,\mathcal{P}\right\rangle
=\sum_{r}g_{r}\left\vert r,\mathcal{P}\right\rangle $. Then the state of the
whole system $\mathcal{PSA}$ after the programming operation reads
\begin{equation}
\left\vert \mathcal{P(SA)}\right\rangle =\sum_{r=0}^{D_{\mathcal{P}}-1}%
g_{r}\left\vert r,\mathcal{P}\right\rangle \left\vert r,\mathcal{SA}%
\right\rangle , \label{psa}%
\end{equation}
where $\left\vert r,\mathcal{SA}\right\rangle =\hat{U}_{\mathcal{SA}}(\hat
{s}_{r},\hat{p}_{r})\left\vert \psi,\mathcal{S}\right\rangle \left\vert
\phi,\mathcal{A}\right\rangle $ encodes the programmed entanglement, if any,
between $\mathcal{S}$ and $\mathcal{A}$. For a given $\left\vert
r,\mathcal{P}\right\rangle $, the pair observables [denoted by ($\hat{s}%
_{r},\hat{p}_{r}$)] and their information to be measured is determined by the
Schmidt form of $\left\vert r,\mathcal{SA}\right\rangle $. Note that
$\left\vert \mathcal{P(SA)}\right\rangle $ can also be written in a Schmidt
form with positive real coefficients \cite{pureEE1}. Hereafter we suppose that
$\left\vert \mathcal{P(SA)}\right\rangle $ has
been Schmidt-decomposed and $g_{r}>0$ (In practice, we could of course have
the situation where $\mathcal{P}$ and $\mathcal{SA}$ are not fully entangled, i.e.,
the number of $g_{r}>0$ is less than $D_{\mathcal{P}}$ or $dD_{\mathcal{A}}$).

Now the key point of our formalism is to require that the programming system
$\mathcal{P}$ encodes all possible, namely, information-complete,
measurement operations that are allowed to act upon the $\mathcal{SA}$-system.
To be \textquotedblleft information-complete\textquotedblright, all
programmed measurement operations $\hat{U}_{\mathcal{SA}}(\hat{s}_{r},\hat
{p}_{r})$ can at least achieve the measurements of a complete set of operators
for $\mathcal{S}$; for the $d$-dimensional system, the complete set has
$d^{2}$\ operators \cite{ThewTomo}, i.e., the minimal $D_{\mathcal{P}}=d^{2}$.
Note that information-complete set of operators or measurement is also
important for quantum state tomography \cite{ICO-Busch,ICM}.

Another trick in the above discussion is that, to enable the information-complete
programmed measurements, it seems that one needs $D_{\mathcal{P}}$
different measurement apparatuses. Hereafter we take a step further by
dropping this specific measurement model by regarding the $\mathcal{A}$-system
as a single quantum system (not necessarily having $\hat{p}$ and $\hat{q}$ as in the
specific model that we considered above) with $D_{\mathcal{A}}$ ($\geq d$)
dimensions and as such, the standard pre-measurement process described above
is simply to entangle $\mathcal{S}$ and $\mathcal{A}$ (see the next Section
for further discussion). In this case we have $D_{\mathcal{P}}\geq
dD_{\mathcal{A}}$ due to the property of the Schmidt decomposition. 
The step is necessary for seeking a model-independent and
intrinsic description of the whole system $\mathcal{PSA}$.

To have an easy understanding of our information-complete description of
physical systems, some remarks are necessary. First, we note that the third
system is also included in other interpretations of QT, such as the
many-worlds interpretation \cite{Everett,Tipler}, the relational
interpretation \cite{Rovelli,Rovelli-book}, and the \textquotedblleft
objective quantum measurement\textquotedblright\ \cite{objectQM}. However, the
third system in our formalism plays a role that is dramatically different from
those interpretations. Actually, imposing information-completeness into our
quantum description of nature distinguishes our theory from all previous
interpretations of QT. Second, the fact that $\left\vert r,\mathcal{SA}%
\right\rangle $, as entangled, can always be written in a Schmidt form implies
a symmetric role played by $\mathcal{S}$ and $\mathcal{A}$. Meanwhile, the
role of $\mathcal{P}$\ is dramatically different from that of either
$\mathcal{S}$ or $\mathcal{A}$. But $\mathcal{P}$ and the combined system
$\mathcal{SA}$ play a symmetric role. We anticipate that such a feature could
have profound consequences, particularly for the internal consistency of the
theory. We will find that this is indeed the case when we consider the
dynamics within the ICQT.

\section{The emergent dual Born rule}

How to acquire information and which kind of information to acquire are two
questions of paramount importance. According to the ICQT, on one hand, the
only way to acquire information is to interact (i.e., entangle) the system
$\mathcal{S}$ and the apparatus $\mathcal{A}$ with each other; no interaction
leads to no entanglement and thus no information. This is in a similar spirit
as the relational interpretation \cite{Rovelli,Rovelli-book}, which treats the
quantum state as being observer-dependent, namely, the state is the relation
between the observer and the system. On the other hand, the programming system
$\mathcal{P}$, by interacting with $\mathcal{SA}$, dictates the way (actually,
the information-complete way) on which kind of information (called
the ``$\mathcal{P}$-$\mathcal{SA}$ information'' hereafter) to acquire about
the system $\mathcal{S}$. For instance, if the whole system is programmed to
measure $\hat{s}_{r}$, then $\mathcal{S}$ and $\mathcal{A}$ interact with each
other to induce the programmed measurement operations $\hat{U}_{\mathcal{SA}%
}(\hat{s}_{r},\hat{p}_{r})$. This process generates the entangled state
$\left\vert r,\mathcal{SA}\right\rangle $ with which $\mathcal{A}$
\textquotedblleft knows\textquotedblright, in a completely coherent way, all
information (called the ``programmed $\mathcal{SA}\left\vert
_{\mathcal{P}}\right.$ information'') about $\mathcal{S}$ in the basis of $\hat{s}_{r}$; the amount of
entanglement contained in $\left\vert r,\mathcal{SA}\right\rangle $ quantifies
the amount of information acquired during this measurement. Also,
$\mathcal{P}$ \textquotedblleft knows\textquotedblright, again in a completely
coherent way, the information about which kind of information (here
$\left\vert r,\mathcal{SA}\right\rangle $) $\mathcal{A}$ has about
$\mathcal{S}$; the amount of the $\mathcal{P}$-$\mathcal{(SA)}$ entanglement
quantifies the amount of information on which kind of measurements to do. All
information is coherently and completely encoded there by certain quantum system.

Thus, for any given system $\mathcal{S}$ in our description one has to ask two
questions: How $\mathcal{S}$ gets entangled with another measurement system
$\mathcal{A}$ and how many independent ways can it be entangled with
$\mathcal{A}$? The answer to the latter question is completely contained in
the entangled state $\left\vert \mathcal{P(SA)}\right\rangle $, while the
answer to the former is the programmed entanglement $\left\vert r,\mathcal{SA}%
\right\rangle $---the two questions are answered by entanglement at two
different levels, called \textit{dual entanglement}.

Now let us state a key point in our ICQT. Namely, \textit{entanglement (created by interaction),
necessary and sufficient for acquiring information, is the measurement and the
physical predictions of the theory} as any possible information (the $\mathcal{P}$-$\mathcal{SA}$ information \emph{and} the programmed $\mathcal{SA}\left\vert
_{\mathcal{P}}\right.$ information) is completely
encoded in the particular dual entanglement structure of the whole system. To
see this, let us note that the reduced density operators for $\mathcal{P}$ and
$\mathcal{SA}$ read
\begin{align}
\rho_{\mathcal{P}}  &  =\mathrm{tr}_{\mathcal{SA}}[\left\vert \mathcal{P(SA)}%
\right\rangle \left\langle \mathcal{P(SA)}\right\vert ]=\sum_{r=0}%
^{D_{\mathcal{P}}-1}g_{r}^{2}\left\vert r,\mathcal{P}\right\rangle
\left\langle r,\mathcal{P}\right\vert ,\nonumber\\
\rho_{\mathcal{SA}}  &  =\mathrm{tr}_{\mathcal{P}}[\left\vert \mathcal{P(SA)}%
\right\rangle \left\langle \mathcal{P(SA)}\right\vert ]=\sum_{r=0}%
^{D_{\mathcal{P}}-1}g_{r}^{2}\left\vert r,\mathcal{SA}\right\rangle
\left\langle r,\mathcal{SA}\right\vert , \label{reduced}%
\end{align}
implying that all information about $\mathcal{P}$ ($\mathcal{SA}$) is
completely contained in the set \{$g_{r}^{2},\left\vert r,\mathcal{P}%
\right\rangle $\} (\{$g_{r}^{2},\left\vert r,\mathcal{SA}\right\rangle $\}),
the physical predictions of the theory. Yet, all these physical predictions
are already encoded completely in the $\mathcal{P}$-$\mathcal{(SA)}$
entanglement. In other words, the $\mathcal{P}$-$\mathcal{(SA)}$ entanglement
is sufficient to predict \{$g_{r}^{2},\left\vert r,\mathcal{P}\right\rangle
$\} and \{$g_{r}^{2},\left\vert r,\mathcal{SA}\right\rangle $\}, a task that
we could expect for a measurement. Similar analysis applies to the programmed
entanglement $\left\vert r,\mathcal{SA}\right\rangle $ as well.

As both $\left\vert r,\mathcal{SA}\right\rangle $ and $\left\vert
\mathcal{P(SA)}\right\rangle $ are pure states, entanglement for each of them is uniquely
quantified by the usual entanglement entropy \cite{pureEE1,pureEE2}. Here, the
$\mathcal{P}$-$\mathcal{(SA)}$ entanglement---a two-party entanglement, although having
three constituents ($\mathcal{P}$, $\mathcal{A}$, and $\mathcal{S}$)---quantifies the $\mathcal{P}$-$\mathcal{SA}$
information and has entanglement entropy maximally $\ln D_{\mathcal{P}}$, while the $\mathcal{S}$-$\mathcal{A}$
entanglement contained in $\left\vert
r,\mathcal{SA}\right\rangle $ quantifies the programmed $\mathcal{SA}\left\vert
_{\mathcal{P}}\right.$ information and has entanglement entropy maximally $\ln D_{\mathcal{S}}$. This
immediately identifies each of the squared coefficients of their Schmidt
decompositions as a probability to reconcile with Shannon's definition of
entropy. Put differently, in our information-complete description of
physical systems, entanglement does be all the information; classical terms
like probability arise in our description because of either our reliance on
classical concept of information or certain approximate and incomplete
description to be shown below. \emph{By regarding entanglement directly as
measurement of complete information, one can avoid the classical-quantum
hybrid feature of current quantum theory or any classical concepts having to
use therein.}

The status of quantum states (more precisely, dual entanglement) in the
information-completeness formalism thus represents a complete reality of the
whole system ($\mathcal{P}$, $\mathcal{S}$, and $\mathcal{A}$, the trinity).
Such a reality picture (\textquotedblleft quantum reality\textquotedblright)
is only possible by taking into account the information-completeness
explicitly in our formalism. Quantum states do exist in a world that is
informational and objective. Whatever an observation might be,
information-complete states always encode information pertaining to that
observation as programmed, without invoking observers or having to appeal to
any mysterious mechanisms to account for wave function collapse; there is
simply no wave function collapse. Here local quantum states (i.e., states for
each of $\mathcal{P}$, $\mathcal{S}$, and $\mathcal{A}$) are all relative, but
information encoded in dual entanglement is invariant under the changes of
local bases, a basic property of entanglement. If one likes, the choice of
local bases can be called a free will or freedom of choice, corresponding to
certain \textquotedblleft gauge\textquotedblright. Yet, all physical
predictions of the theory are encoded in dual entanglement and do not depend
on the chosen gauge. To compare with the relational
interpretation \cite{Rovelli,Rovelli-book}, the ICQT tells us a kind of \emph{quantum
relationalism}---While local quantum states of a single system are merely of relative meaning,
the relations between states of two systems have physical significance; in a truly quantum
world, entanglement is \emph{the} relations.

In certain sense, $\mathcal{P}$ and $\mathcal{A}$ act like a \textquotedblleft
quantum being\textquotedblright\ (\textquotedblleft qubeing\textquotedblright)
who holds coherently all the information-complete programmes on how to
entangle $\mathcal{S}$ and $\mathcal{A}$. In this way, the qubeing has all the
information about $\mathcal{S}$. However, our human beings plus the measurement
apparatuses, unlike the qubeing, are macroscopic and have so many quantum degrees of freedom.
For example, an experimenter, Alice, together with her apparatus, would like to acquire
information about $\left\vert \psi,\mathcal{S}\right\rangle $. First of all,
she has to \textit{decide} which kind of information she would like to know.
After making a decision, she needs then to \textit{observe} (that is, to
interact with) her apparatus readily entangled with $\mathcal{S}$. In
principle, Alice's decisions and observations are all physical processes which
should be described quantum mechanically. Thus, if the total system is isolated,
it must be in dual entanglement encoding complete information of the whole system,
as argued above. Nevertheless, \emph{Alice, a human being, is so used to and familiar with
classical concepts on information and physical systems}. Practically, she has
limited ability and is lack of full knowledge of the entire system. In this case,
she has to \textquotedblleft trace out\textquotedblright\ those quantum
degrees of freedom involved in her decisions (interaction between $\mathcal{P}$ and $\mathcal{SA}$),
leading to a mixed state $\sum_{r}\left\vert g_{r}\right\vert ^{2}\left\vert
r,\mathcal{P}\right\rangle \vert \left\langle r,\mathcal{P}\right\vert $ and
$\sum_{r}\left\vert g_{r}\right\vert ^{2}\left\vert
r,\mathcal{SA}\right\rangle \vert \left\langle r,\mathcal{SA}\right\vert $ [see Eq.~(\ref{reduced})]. This state allows a probability interpretation about Alice's freedom of choice: Each of
her decisions $\left\vert r,\mathcal{P}\right\rangle $ occurs with a
probability of $\left\vert g_{r}\right\vert ^{2}$. As far as a particular
choice $\left\vert r,\mathcal{P}\right\rangle $ has been made, again she has
to trace out her quantum degrees of freedom involved in her observation
(interaction between $\mathcal{A}$ and $\mathcal{S}$). This then leads to the usual Born rule about
$\left\vert \psi,\mathcal{S}\right\rangle $ for the given measurement. Thus,
in the ICQT, the Born rule, also in dual form, is an emergent or derived rule
determined by the dual entanglement structure.

To summarize the above picture, the world view of the ICQT is fascinating. If
we regard the system $\mathcal{S}$ as an indivisible part of the qubeing
$\mathcal{PA}$, the whole system $\mathcal{PSA}$ then represents an
information-complete and objective entity; it seems that the qubeing has
its own \textquotedblleft consciousness\textquotedblright, a kind of
miraculous quantum ability, to encode and access all its information in the
form of dual entanglement, in which the constituent parts of the qubeing are
mutually measured or defined. In other words, for the qubeing all information
(namely, all physical predictions) is encoded in dual entanglement via
interaction, but not obtained via the usual quantum measurement with the
unavoidable concept of the wave function collapse.

\begin{figure}[ptb]
\centerline{\includegraphics[width=.40\textwidth]{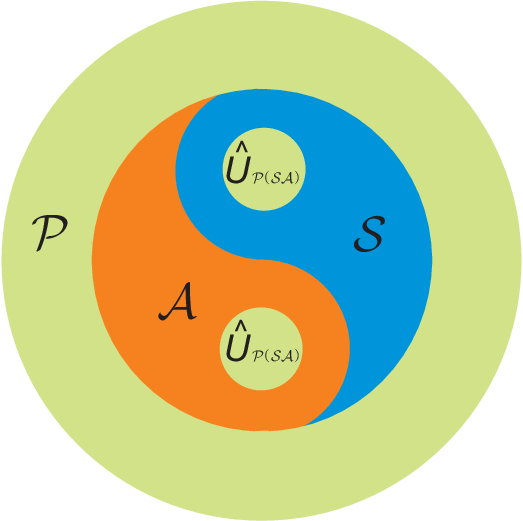}}\caption{The
trinary picture of the world. The division of system $\mathcal{S}$,
measurement apparatus $\mathcal{A}$, and programming system $\mathcal{P}$
naturally arises in the information-complete description of physical
systems. \textquotedblleft The Taiji pattern\textquotedblright\ shows in an
intuitive manner the $\mathcal{S}$-$\mathcal{A}$ interaction (entanglement),
while the green discs inside and outside the Taiji pattern represent the
programmed measurement operations $\hat{U}_{\mathcal{P(SA)}}$ between
$\mathcal{P}$ and $\mathcal{SA}$. In ancient China, Taoists regarded the Taiji
pattern as a \textquotedblleft diagram of the Universe\textquotedblright. The
trinary picture of the world shown here is ubiquitous in the sense that the
world, at the most fundamental level, is made up of a trinity: gravity (i.e.,
spacetime, $\mathcal{P}$), elementary matter fermions ($\mathcal{S}$) and
their gauge fields ($\mathcal{A}$); the trinity should be describable by the
ICQT.}%
\label{TJ}%
\end{figure}

\emph{The trinary picture (the division of $\mathcal{S}$, $\mathcal{A}$, and
$\mathcal{P}$) of physical systems arises here as a new feature of the ICQT},
as shown in Fig.~1. To retain the information-complete description of
nature, such a trinary picture seems to be unavoidable. The limitation of
information contents in dual entanglement could be tentatively called
\textquotedblleft the trinity principle\textquotedblright, instead of the
conventional complementarity principle, to put the trinary property of
physical systems on the most fundamental ground.

The loss of the trinary picture of describing physical systems leads to the
\textit{emergent dual Born rule}, i.e., the probability description on which
kind of observables to measure and then on which eigenvalue of the observable
to measure, due to, e.g., lack of full knowledge of the entire system in our
ICQT. The conventional von Neumann entropy quantifies this dual loss of
information. In other words, the conventional Born rule arises as a
consequence of the sacrifice of information-complete description in the
trinary picture; the sacrifice leads to a \textit{partial} reality of physical
system as described by conventional QT.

\section{The information-complete dynamics}

According to the above picture of nature, single free systems are simply
meaningless for acquiring information; a system, which does not give
information to (i.e., interact with) other systems in any way, simply does not
exist. The \textquotedblleft$\mathcal{S}$+$\mathcal{A}$\textquotedblright%
\ description in the usual QT is also inadequate because of its information-incompleteness as there is certain information (e.g., the choice of measurement bases)
not encoded by any physical quantum system. Therefore, the dynamics of the ICQT will be dramatically
different from the usual picture as it requires interacting $\mathcal{P}%
$+$\mathcal{SA}$ so as to obey the information-completeness principle within
the trinary picture. Without specifying the $\mathcal{P}$+$\mathcal{SA}$
dynamics to maintain the information-complete trinary description, it is
meaningless or information-incomplete to specify local states of single
systems in $\mathcal{P}$+$\mathcal{SA}$. Namely, \textit{the ICQT is
characterized by the indivisibility of its kinematics and dynamics}. Below, we
give some key features of the information-complete dynamics.

Before considering the information-complete dynamics, let us introduce an
important concept of \textit{dual} \textit{measurability}: the $\mathcal{P}%
$-$\mathcal{SA}$ measurability and the programmed $\mathcal{SA}\left\vert
_{\mathcal{P}}\right.  $ measurability. The former means the ability of
measuring $\mathcal{P}$ with $\mathcal{SA}$ and vice versa; the latter means
the ability of measuring $\mathcal{A}$ with $\mathcal{S}$ and vice versa,
under a given programmed measurement operation of $\mathcal{P}$. The
$\mathcal{P}$-$\mathcal{SA}$ measurability (the programmed measurability
$\mathcal{SA}\left\vert _{\mathcal{P}}\right.  $) leads to $D_{\mathcal{P}%
}=D_{\mathcal{A}}D_{\mathcal{S}}$ ($D_{\mathcal{A}}=D_{\mathcal{S}}=D$) and
thus a symmetric role between $\mathcal{P}$ and $\mathcal{SA}$ ($\mathcal{S}$
and $\mathcal{A}$). Note that here measurability does not mean certain quantum
measurement actually performed by an experimenter in the usual sense.
As we pointed out above, in the ICQT entanglement is the measurement. Thus,
dual measurability is simply another side of dual entanglement; the relation among
the dimensions for the trinity's quantum states is then a simple consequence of
the Schmidt decompositions of dual entanglement, which is the complete physical prediction
of our theory.

After the above preparation, now we give a definition of
information-complete physical systems: A physical system is said to be
information-complete if and only if (the use of \textquotedblleft if and
only if\textquotedblright\ will be explained below) it is consisted of
$\mathcal{S}$, $\mathcal{A}$ and $\mathcal{P}$ described as a trinity such
that the $\mathcal{P}$-$\mathcal{SA}$ measurability and the programmed
measurability $\mathcal{SA}\left\vert _{\mathcal{P}}\right.  $ are satisfied.
As a result of this definition, the $\mathcal{P}$-$\mathcal{SA}$ measurability
implies the existence of at most $D^{2}$ independent \textit{information-complete
measurement (entanglement) operations} in the Hilbert spaces of both
$\mathcal{P}$ and $\mathcal{SA}$; these operations generate at most $D^{2}%
$-dimensional entanglement between $\mathcal{P}$ and $\mathcal{SA}$ and at
most $D^{2}$ entangled states between $\mathcal{S}$ and $\mathcal{A}$ such
that $\mathcal{P}$ and $\mathcal{SA}$ are mutually measuring and defining.
Meanwhile, under the given programming state of $\mathcal{P}$, the programmed
measurability $\mathcal{SA}\left\vert _{\mathcal{P}}\right.  $ implies the
existence of at most $D$-dimensional entangled states between $\mathcal{S}$
and $\mathcal{A}$ such that $\mathcal{S}$ and $\mathcal{A}$ are mutually
measuring and defining, as programmed. For convenience, we also call
\textit{the complete set of states and operators defined in the Hilbert space
of }$P$\textit{ or }$SA$\textit{ are information-complete. Accordingly,
states and operators for either }$S$\textit{ or }$A$\textit{ alone are
information-incomplete. }Thus, in the ICQT the role of observables defined
for either $\mathcal{S}$ or $\mathcal{A}$ is quite different from the role of
observables defined for $\mathcal{P}$.

Let us suppose that the information-complete system $\mathcal{PSA}$ has a
general Hamiltonian $\hat{H}_{\mathcal{PSA}}$. We assume that the whole system
evolutes according to the standard Schr\"{o}dinger equation (we take $\hbar=1$),
namely, $i\frac{d}{dt}\left\vert \mathcal{P(SA}),t\right\rangle =\hat
{H}_{\mathcal{PSA}}\left\vert \mathcal{PSA},t\right\rangle $. In general,
$\hat{H}_{\mathcal{PSA}}=\hat{H}_{\mathcal{P}}+\hat{H}_{\mathcal{S}}+\hat
{H}_{\mathcal{A}}+\hat{H}_{\mathcal{PS}}+\hat{H}_{\mathcal{PA}}+\hat
{H}_{\mathcal{SA}}+\hat{H}_{\mathcal{P(SA)}}$, where the subscripts label the
corresponding systems. Now our problem is to determine how the information-completeness
principle constrains the form of $\hat{H}_{\mathcal{PSA}}$ and
thus the dynamics of the $\mathcal{PSA}$-system.

Note that we can choose an orthonormal basis (the \textquotedblleft
programming basis\textquotedblright) \{$\left\vert e_{n},\mathcal{P}%
\right\rangle ;n=0,1,..D_{\mathcal{P}}-1$\} to span the whole Hilbert space of
$\mathcal{P}$. We associate each programming state $\left\vert e_{n}%
,\mathcal{P}\right\rangle $ as an eigenstate of $\mathcal{P}$-system's
\textquotedblleft programming observable\textquotedblright\ $\hat
{e}_{\mathcal{P}}$ with eigenvalue $e_{n}$. It is easy to verify that the
following Hamiltonian obeys the informational completeness principle:
\begin{align}
\mathcal{\hat{I}}_{\mathcal{PSA}}  &  =\hat{H}_{\mathcal{P}}+\hat
{H}_{\mathcal{P(SA)}}=\hat{H}_{\mathcal{P}}\nonumber\\
&  +\sum_{n=0}^{D_{\mathcal{P}}-1}\left\vert e_{n},\mathcal{P}\right\rangle
\left\langle e_{n},\mathcal{P}\right\vert \hat{H}_{\mathcal{SA}\left\vert
_{\mathcal{P}}\right.  }(e_{n},t), \label{IPSA}%
\end{align}
if we impose the $\mathcal{P}$-$\mathcal{SA}$ measurability condition as
follows
\begin{equation}
\lbrack\hat{H}_{\mathcal{P(SA)}},\hat{e}_{\mathcal{P}}]=[\hat{H}_{\mathcal{P}%
},\hat{e}_{\mathcal{P}}]=0, \label{pmc}%
\end{equation}
namely, $\hat{e}_{\mathcal{P}}$ commutes with both $\hat{H}_{\mathcal{P(SA)}}$
and $\hat{H}_{\mathcal{P}}$. If one takes $\hat{e}_{\mathcal{P}}=\hat
{H}_{\mathcal{P}}$, Eq.~(\ref{pmc}) reads simply as
\begin{equation}
\lbrack\hat{H}_{\mathcal{P(SA)}},\hat{H}_{\mathcal{P}}]=0. \label{hhp}%
\end{equation}

In Eq.~(\ref{IPSA}) $\hat{H}_{\mathcal{SA}\left\vert _{\mathcal{P}}\right.
}(e_{n},t)=i\frac{d}{dt}\hat{U}_{\mathcal{SA}\left\vert _{\mathcal{P}}\right.
}\left(  e_{n},t\right)  \cdot\hat{U}_{\mathcal{SA}\left\vert _{\mathcal{P}%
}\right.  }^{-1}\left(  e_{n},t\right)  $ is an information-complete
operator set of $\mathcal{SA}$; $\sum_{n=0}^{D_{\mathcal{P}}-1}\left\vert e_{n},\mathcal{P}\right\rangle
\left\langle e_{n},\mathcal{P}\right\vert \hat{H}_{\mathcal{SA}\left\vert
_{\mathcal{P}}\right.  }(e_{n},t)$ represents actually the spectrum-decomposition with respect to $\hat{e}_{\mathcal{P}}$ of $\hat
{H}_{\mathcal{P(SA)}}(\hat{e}_{\mathcal{P}},t)$. Meanwhile, $\mathcal{\hat{I}}_{\mathcal{PSA}}$
takes a form like $\mathcal{\hat{I}}_{\mathcal{PSA}}=\hat{H}_{\mathcal{P}%
}+\hat{H}_{\mathcal{PS}}+\hat{H}_{\mathcal{PA}}+\hat{H}_{\mathcal{P(SA)}}$
such that $\mathcal{P}$ \textit{universally} couples with $\mathcal{S}$ and
$\mathcal{A}$. Note that in $\mathcal{\hat{I}}_{\mathcal{PSA}}$, Hamiltonians $\hat
{H}_{\mathcal{S}}$, $\hat{H}_{\mathcal{A}}$, and $\hat{H}_{\mathcal{SA}}$ do
not appear in $\mathcal{\hat{I}}_{\mathcal{PSA}}$. These \textquotedblleft
local\textquotedblright\ Hamiltonians ($\hat{H}_{\mathcal{S}}$, $\hat
{H}_{\mathcal{A}}$, and $\hat{H}_{\mathcal{SA}}$) decoupled from $\mathcal{P}$
induce local unitary transformations upon $\mathcal{S}$ or/and $\mathcal{A}$,
corresponding to certain gauge, which can be gauged out in physical
predictions encoded in dual entanglement.

$\mathcal{\hat{I}}_{\mathcal{PSA}}$ induces an evolution $\left\vert
\mathcal{P(SA}),t\right\rangle =\hat{U}_{\mathcal{PSA}}\left\vert
\mathcal{P},t=0\right\rangle \left\vert \mathcal{SA},t=0\right\rangle $; the
evolution operator $\hat{U}_{\mathcal{PSA}}$ always has a factorizable
structure
\begin{equation}
\hat{U}_{\mathcal{PSA}}(t)=\sum_{n=0}^{D_{\mathcal{P}}-1}\left\vert
e_{n},\mathcal{P}\right\rangle \left\langle e_{n},\mathcal{P}\right\vert
\hat{U}_{\mathcal{P}}(t)\hat{U}_{\mathcal{SA}\left\vert _{\mathcal{P}}\right.
}(e_{n},t), \label{factor}%
\end{equation}
as a result of the $\mathcal{P}$-$\mathcal{SA}$ measurability condition
(\ref{pmc}), such that $(\forall e_{n})$
\begin{align}
i\frac{d}{dt}\hat{U}_{\mathcal{P}}(t)  &  =\hat{H}_{\mathcal{P}}\hat
{U}_{\mathcal{P}}(t),\nonumber\\
i\frac{d}{dt}\hat{U}_{\mathcal{SA}\left\vert _{\mathcal{P}}\right.  }%
(e_{n},t)  &  =\hat{H}_{\mathcal{SA}\left\vert _{\mathcal{P}}\right.  }%
(e_{n},t)\hat{U}_{\mathcal{SA}\left\vert _{\mathcal{P}}\right.  }(e_{n},t).
\label{dual-dyn}%
\end{align}
In this way, the dynamical evolutions of $\mathcal{P}$ and $\mathcal{SA}$ are
mutually defined, in accordance with the information-completeness principle.

The Hamiltonian $\mathcal{\hat{I}}_{\mathcal{PSA}}$ as given above respects
the $\mathcal{P}$-$\mathcal{SA}$ measurability. If we also require the
programmed measurability $\mathcal{SA}\left\vert _{\mathcal{P}}\right.  $
($\forall e_{n}$), the evolution governed by $\hat{H}_{\mathcal{SA}\left\vert
_{\mathcal{P}}\right.  }(e_{n},t)$ depends on which system ($\mathcal{S}$ or
$\mathcal{A}$) defines the programming observable. For example, if one can
find the programming observable $\hat{\varepsilon}_{\mathcal{S}\left\vert
_{\mathcal{P}}\right.  }$ for $\mathcal{S}$, the programmed evolution for
$\mathcal{SA}$\ will similarly acquire the factorizable structure as
\begin{align}
\hat{H}_{\mathcal{SA}\left\vert _{\mathcal{P}}\right.  }(e_{n},t) &
=\sum_{i=0}^{D-1}\left\vert \varepsilon_{i}(e_{n}),e_{n},\mathcal{S}%
\right\rangle \left\langle \varepsilon_{i}(e_{n}),e_{n},\mathcal{S}\right\vert
\nonumber\\
&  \times\hat{H}_{\mathcal{A}_{\left\vert S\right.  }\left\vert _{\mathcal{P}%
}\right.  }\left[  e_{n},\varepsilon_{i}(e_{n}),t\right]  \nonumber\\
&  +\hat{H}_{\mathcal{S}\left\vert _{\mathcal{P}}\right.  }(e_{n},t)+\hat
{H}_{\mathcal{A}\left\vert _{\mathcal{P}}\right.  }(e_{n},t)\label{Hsap}%
\end{align}
with $\hat{H}_{\mathcal{A}_{\left\vert S\right.  }\left\vert _{\mathcal{P}%
}\right.  }\equiv i\frac{d}{dt}\hat{U}_{\mathcal{A}_{\left\vert S\right.
}\left\vert _{\mathcal{P}}\right.  }\left[  e_{n},\varepsilon_{i}%
(e_{n}),t\right]  \cdot\hat{U}_{\mathcal{A}_{\left\vert S\right.  }\left\vert
_{\mathcal{P}}\right.  }^{-1}\left[  e_{n},\varepsilon_{i}(e_{n}),t\right]  $.
Here the orthonormal basis for $\mathcal{S}\left\vert _{\mathcal{P}}\right.  $
is \{$\left\vert \varepsilon_{i}(e_{n}),e_{n},\mathcal{S}\right\rangle $\},
where $\left\vert \varepsilon_{i}(e_{n}),e_{n},\mathcal{S}\right\rangle $ is
an eigenstate of $\hat{\varepsilon}_{\mathcal{S}\left\vert _{\mathcal{P}%
}\right.  }(e_{n})$ with eigenvalue $\varepsilon_{i}(e_{n})$ for given $e_{n}%
$. Similarly to the $\mathcal{P}$-$\mathcal{SA}$ measurability condition
(\ref{pmc}), we need to impose the programmed measurability $\mathcal{SA}%
\left\vert _{\mathcal{P}}\right.  $ condition $(\forall e_{n})$
\begin{equation}
\lbrack\hat{H}_{\mathcal{SA}\left\vert _{\mathcal{P}}\right.  }(e_{n}%
,t),\hat{\varepsilon}_{\mathcal{S}\left\vert _{\mathcal{P}}\right.  }%
(e_{n})]=[\hat{\varepsilon}_{\mathcal{S}\left\vert _{\mathcal{P}}\right.
}(e_{n}),\hat{H}_{\mathcal{S}\left\vert _{\mathcal{P}}\right.  }%
(e_{n},t)]=0.\label{sapmc}%
\end{equation}
The $\mathcal{SA}\left\vert _{\mathcal{P}}\right.  $ dynamics is then similar
to the $\mathcal{P}$-$\mathcal{SA}$ dynamics considered above. Such a \textit{dual dynamics} of the whole system $\mathcal{PSA}$ is an attribute of
the trinary description and quite distinct to the usual Schr\"{o}dinger evolution.

What is the physical significance of the programming basis \{$\left\vert
e_{n},\mathcal{P}\right\rangle $\} and the associated observable $\hat
{e}_{\mathcal{P}}$? Actually it is physically transparent that \{$\left\vert
e_{n},\mathcal{P}\right\rangle $\} as the physical predictions can be
identified with the Schmidt basis for the $\mathcal{P}$-$\mathcal{SA}$
decomposition, which is not affected by the local transformations generated by
$\hat{H}_{\mathcal{P}}$. A more interesting possibility is to interpret
$\hat{e}_{\mathcal{P}}$ as a quantum nondemolition observable
\cite{QND-science,QND-rmp}; note that the pre-measurement involves a
nondemolition coupling between $\mathcal{S}$ and $\mathcal{A}$ \cite{objectQM}%
. Then the $\mathcal{P}$-$\mathcal{SA}$ measurability condition (\ref{pmc}) is
a (sufficient) condition for a quantum nondemolition measurement of $\hat
{e}_{\mathcal{P}}$. In the context of quantum nondemolition observable, a more
general measurability condition could be imposed. For instance, the
$\mathcal{P}$-$\mathcal{SA}$ measurability condition in Eq.~(\ref{pmc}) might
be replaced by
\begin{equation}
\lbrack\hat{U}_{\mathcal{P}}(t),\hat{e}_{\mathcal{P}}]\left\vert
\mathcal{P(SA}),t\right\rangle =[\hat{U}_{\mathcal{PSA}}(t),\hat
{e}_{\mathcal{P}}]\left\vert \mathcal{P(SA}),t\right\rangle =0. \label{UeUe}%
\end{equation}

As we require that the evolution $\left\vert \mathcal{P(SA}),t\right\rangle
=\hat{U}_{\mathcal{PSA}}\left\vert \mathcal{P},t=0\right\rangle \left\vert
\mathcal{SA},t=0\right\rangle $ results in a state already in the Schmidt
form
\begin{align}
\left\vert \mathcal{P(SA)},t\right\rangle  &  =\sum_{r=0}^{D_{\mathcal{P}}%
-1}g_{r}\left\vert e_{n},\mathcal{P}\right\rangle \left\vert e_{n}%
,\mathcal{SA}\right\rangle ,\nonumber\\
\left\vert e_{n},\mathcal{SA}\right\rangle  &  \equiv\hat{U}_{\mathcal{SA}%
\left\vert _{\mathcal{P}}\right.  }(e_{n},t)\left\vert \mathcal{SA}%
,t=0\right\rangle , \label{PSASchm}%
\end{align}
it is easy to prove that \{$\left\vert e_{n},\mathcal{SA}\right\rangle $\}
forms an orthonormal basis [This is true, e.g., if $\left\vert e_{n}%
,\mathcal{SA}\right\rangle $ is the eigenstate of $\hat{H}_{\mathcal{SA}%
\left\vert _{\mathcal{P}}\right.  }(e_{n},t)$ with eigenvalue $E_{\mathcal{SA}%
\left\vert _{\mathcal{P}}\right.  }(e_{n},t)$] and
\begin{equation}
\hat{U}_{\mathcal{P}}(t)\left\vert \mathcal{P},t=0\right\rangle =\sum
_{r=0}^{D_{\mathcal{P}}-1}g_{r}\left\vert e_{n},\mathcal{P}\right\rangle .
\label{Pstate}%
\end{equation}
Equation (\ref{Pstate}) looks as if the single system $\mathcal{P}$ brings the
properties \{$g_{r}^{2},\left\vert e_{n},\mathcal{P}\right\rangle $\} of the
whole $\mathcal{P}$-$\mathcal{SA}$ system. This is of course not true as the
Schmidt basis is the joint properties of the whole system. The specific form
of $\hat{U}_{\mathcal{P}}(t)\left\vert \mathcal{P},t=0\right\rangle $ stems
from the specific choice of $\hat{e}_{\mathcal{P}}$ in the Schmidt basis of
$\left\vert \mathcal{P(SA)},t\right\rangle $. Actually, in this case we can
choose $\hat{e}_{\mathcal{P}}=\rho_{\mathcal{P}}$.

When applying the ICQT to quantum gravity coupled with matter quantum fields
\cite{icQFT}, the usual Schr\"{o}dinger equation losses its dynamical meaning
and becomes trivially a constraint $\hat
{H}_{\mathcal{PSA}}\left\vert \mathcal{P(SA})\right\rangle =0$, known as the
Wheeler-DeWitt equation. In this case, the programming observable $\hat
{e}_{\mathcal{P}}$ then corresponds the Dirac observable
\cite{Rovelli-book,Thiemann}; dynamics described by a pair of the Schr\"{o}dinger equations
can be recovered due to the dual entanglement structure. Therefore, \textit{in this field-theoretical case, the significance and necessity of the information-complete dynamics in trinity
is physically more transparent}. Moreover, if $\mathcal{P}$ is the
quantized gravitational field, then the \ Hamiltonians $\hat
{H}_{\mathcal{S}}$, $\hat{H}_{\mathcal{A}}$, and $\hat{H}_{\mathcal{SA}}$ are
simply ruled out as the gravitational field universally couples with any form
of matter.

To end this Section, it is important to note that the distinguished roles of
$\mathcal{P}$ and $\mathcal{SA}$ are relative. Depending on the specific form
of the trinary Hamiltonian, we could have another possibility that
$\mathcal{SA}$ can programme the evolution of $\mathcal{P}$. In this case, the
programming basis is chosen for $\mathcal{SA}$ and associated with an
observable being commutative with $\hat{H}_{\mathcal{SA}}$ such that the
$\mathcal{P}$-$\mathcal{SA}$ measurability and a similar dynamics as in
Eq.~(\ref{dual-dyn}) can still be obtained. Furthermore, the roles of
$\mathcal{P}$ and $\mathcal{SA}$ are actually symmetric due to a nice property
of the Schmidt decomposition \cite{Schmidt}, in which if an orthonormal basis
labelled by an index (e.g., $n$) is chosen for a system, then the orthonormal
basis for another system, by acting a unitary transformation upon it, is
labelled by the same index. This property implies that both of the bases are
already the Schmidt bases, up to local unitary transformations. A similar
consideration is applicable to the programmed measurability $\mathcal{SA}%
\left\vert _{\mathcal{P}}\right.  $, too.

\section{Relation with conventional quantum theory}

What is the relation between the ICQT and the usual QT? Before answering this
question, first of all we have to ask ourselves: Why are we bothered to revise
the conventional QT into the current formulation of such a strange appearance?
Here we must introduce at the very beginning the interacting/entangling
$\mathcal{P}$+$\mathcal{SA}$ trinity with a dynamical evolution (being always
unitary) determined by the information-completeness principle such that the
physical properties of, e.g., $\mathcal{P}$ and $\mathcal{SA}$ are coherently
and completely stored in the $\mathcal{P}$-$\mathcal{SA}$ entanglement and can
only be predicted conditionally on each other. Anyway, in non-relativistic
quantum mechanics there seems to be no physical motivation to introduce
$\mathcal{P}$. This is in sharp contrast to traditional QT, where the
\textquotedblleft$\mathcal{S}$+$\mathcal{A}$\textquotedblright\ description is
sufficient and isolated, single systems (free particles, free quantum fields,
and so on) can have certain physical properties which can be accessed by a
mysterious and non-unitary measurement process.

However, one of the most
important lessons learned from general relativity is that spacetime is
dynamical and the same thing as gravity. Therefore, \emph{there are, even in
principle, no perfectly isolated systems as they must live in and couple with
dynamical spacetime}. If we quantize everything of nature, even gravity
(spacetime), which mechanism could trigger a non-unitary measurement process?
Of course, one could simply ignore, as an approximation, the dynamical and
quantum nature of spacetime as the common wisdom does. Then, why such an
ignorance could be a safe approximation without causing any internal
inconsistency or incompleteness of traditional QT? In any case, spacetime is
such an elementary physical entity. Anyone who does not shut eyes to these
problems, among other interpretational difficulties, has to conclude that
\emph{traditional QT must be incomplete as far as spacetime is not treated physically
as a quantum system}; the information-completeness principle is a possible remedy to complete current
quantum formalism, as we suggest.

Ultimately, we should describe nature with quantum field theory. An
information-complete quantum field theory \cite{icQFT} can indeed be
formulated; the ICQT developed here for finite-dimensional quantum mechanical
systems is thus the conceptual preparation and the mathematical formulation
for that purpose. Therein, if we regard $\mathcal{S}$ as particle (i.e.,
matter fermion) fields and $\mathcal{A}$ as their gauge fields, and
$\mathcal{SA}$ together as matter fields, then we immediately recognize that
system $\mathcal{P}$ must be the gravitational field (i.e., spacetime),
nothing else, as only gravitational field, while self-interacting, universally
interacts with all other fields. Recall that in our $\mathcal{P}%
$+$\mathcal{SA}$ trinity, $\mathcal{P}$ must universally couple with
$\mathcal{S}$ and $\mathcal{A}$. If we think this way, an amazing picture
(Fig.~1) of our world arises: The gravitational field and matter fields are
mutually defined and entangled---no matter, no gravity (spacetime) and vice
versa, and for each of their entangled patterns, matter fermion fields and
their gauge fields are likewise mutually defined and entangled. If this is
indeed what our nature works to obey the information-completeness principle
guaranteeing the completeness of the theory from the outset, the conventional
QT will be an approximation of our ICQT when we ignore quantum effects of
nature's programming system, i.e., gravity. Under such an approximation the
ICQT reduces to conventional QT, characterized by the usual Schr\"{o}dinger
equation and the probability description, and as such, QT in its current form
is thus information-incomplete. This is in the exact sense that classical
Newtonian mechanics is an approximate theory of special relativity when a
physical system has a speed much less than the speed of light.

On the other hand, no matter how weak gravity is, it is forced to be there by
the information-completeness principle, to play a role for completing a
consistent quantum theory. This unique role of gravity (or spacetime) in our
theory is consistent with the remarkable fact that only gravity is universally
coupled to all other physical fields (matter fermion fields and their gauge fields). Of
course, our current quantum description is an extremely good approximation.
But for scales near the Planck one and for early Universe, quantized spacetime
acts as the programming system and the ICQT will be necessary. Thus, both
facts (i.e., current QT works so well and quantum gravity effects are so weak
at normal scales) hide so deeply any new theoretical architecture beyond
current QT, like the ICQT. Even in the string theory and loop quantum gravity, there is no change of
the underlying quantum structure. By contrast, what we suggest within the ICQT
is that at the level of quantized fields including quantized spacetime,
everything is quantized and one does not have the usual separation of quantum
systems and observers. In this case, one has to give up the classical-quantum
hybrid feature of current QT. For this purpose, the most obvious way seems to
be the elimination of the measurement postulate in our fully quantum (namely,
not classical-quantum hybrid) description of nature. As we hope to argue,
giving up the classical concepts associated with the measurement postulate in
current QT does not lead to any sacrifice of our predictive power as the
complete information is encoded by the dual entanglement structure.

If we take the above argument seriously, then \textit{the ICQT captures the
most remarkable trinity of nature}, namely, the division of nature by matter
fermions, their gauge fields, and gravity (spacetime), though the role of the
Higgs field needs a separate consideration (see Ref.~\cite{gGUT} on this issue). The previous two sections argued the necessity of the information-completeness in the trinary description.
Here we see that it is also sufficient: We do not have to invoke more
programming systems to program $\mathcal{PSA}$ simply because \textit{we do
not have spacetime (gravity) out of spacetime (gravity)---Trinity is necessary
and sufficient}. This eliminates the von Neumann chain in the usual quantum
measurement model.

One of the most challenging problems in current physics is how to put QT and
general relativity into a single, consistent theory. To achieve this, it is
encouraging to have a quantum formalism like the ICQT, in which gravity must
be quantized and plays an essential role. As we showed elsewhere \cite{icQFT},
following the above arguments indeed leads to a consistent quantum framework
of unifying spacetime (gravity) and matter, without the fundamental
inconsistencies \cite{Thiemann} between gravity and conventional quantum field
theory, implying the conceptual advantages of our theory. For instance, with
the theoretical input from loop quantum gravity predicting the quantized
geometry \cite{Rovelli-book,Thiemann,quanA1,quanA2,quanA3}, the
information-complete quantum field theory naturally explains the holographic
principle, as well as its generalization, via spacetime-matter entanglement.
As we currently understand it, the holographic
principle \cite{tHooft,Susskind,holoRMP} imposes a strong limit on the allowed states of quantum system in any finite spacetime regime. Such a limit paves the way to escape the infrared and ultraviolet singularities
(divergences) that occur in conventional quantum field theory \cite{Rovelli-book,Thiemann}.

Thus, the ICQT gives a strong motivation or reason for quantizing
spacetime/gravity; there is no trinity if there is no quantized gravity. The
natural position of gravity in the ICQT cannot be accidental and may be a
strong evidence supporting our information-complete description of nature.
It is surprise to see that nature singles out gravity as a programming field,
which plays a role that is definitely different from matter fields. However,
quantizing gravity as yet another field, as in conventional quantum field
theory, is not sufficient and does not automatically result in a correct and
consistent quantum theory of all known forces. Only when the information-completeness
in the trinary description is integrated into our quantum
formulation, can we have the desired theory of the Universe. The distinct
roles of matter-matter (fermions and their gauge fields) entanglement and
spacetime-matter (i.e., gravity-matter) entanglement indicate the reason why
quantizing gravity as usual quantized fields suffers from well-known
conceptual problems.

As an abstract mathematical structure, current QT is content-irrelevant in the
following sense. While it is believed to be universally applicable to physical
systems of any physical contents, ranging from elementary particles and
(super)strings to the whole Universe, what physical content that it describes
does not matter and the physical content never changes its very structure. The
situation for classical mechanics is quite similar in this aspect. However,
the ICQT changes this in a dramatic way in the sense that the trinary picture
of nature has to be integrated into a consistent formulation to enable an
information-complete description. The physical content that the ICQT
describes does matter as the states and their dynamical evolution of the
trinary system are constrained or structured into the dual forms specified
above. In particular, the inclusion of the programming system, identified with
gravity in the field-theoretical case, is very essential and necessary in our description.

\section{Information-complete quantum computation}

A new theory should make new predictions or/and motivate new applications; for new predictions of our
theory, see Refs.~\cite{icQFT,gGUT}. Of
course, previous interpretations of QT are very important for a better
understanding of the theory. However, no interpretations make new predictions
or/and motivate fundamentally new applications. Now we argue that our ICQT
indeed motivates new applications if we consider its computational power. Even
though gravity would play certain role in our future understanding of nature,
artificial information-complete quantum systems are realizable without quantizing gravity.

What is an information-complete quantum computer (ICQC)? We define the
ICQC as an artificial information-complete quantum systems, or a quantum
intelligent system (qubeing), which has an information-complete trinary
structure consisting of $\mathcal{S}$, $\mathcal{A}$, and $\mathcal{P}$. The
ICQC starts from an initial state $\left\vert \mathrm{IC}\text{\textrm{QC}%
}\right\rangle _{0}=\left\vert \psi,\mathcal{S}\right\rangle \left\vert
\phi,\mathcal{A}\right\rangle \left\vert \chi,\mathcal{P}\right\rangle $. As
usual, the $\mathcal{S}$ system has $n$ qubits, and thus dimensions of $2^{n}%
$. To be well defined, we also use qubits to make up the $\mathcal{A}$ system
and the $\mathcal{P}$ system; $\mathcal{A}$ ($\mathcal{P}$) has
$n_{\mathcal{A}}$ ($n_{\mathcal{P}}$) qubits and dimensions of
$2^{n_{\mathcal{A}}}$ ($2^{n_{\mathcal{P}}}$). To satisfy the information-completeness
principle, we have $n_{\mathcal{A}}=n$\ and $n_{\mathcal{P}}=2n$.
Our ICQC then works by applying certain patterns of quantum logic
gates (single-qubit, two-qubit, and three-qubit gates), determined by quantum algorithm
pertaining to the question under study. Generally speaking, as an artificially
controllable quantum system the patterns of gates are allowed to exhaust all
unitary operations on the whole $\mathcal{PSA}$ system. Here, for our purpose
we consider a simplified ICQC, namely, we only perform the programmed
measurement operation $\hat{U}_{\mathcal{P(SA)}}$ on $\mathcal{PSA}$. The
resulting final state of the ICQC reads $\left\vert \mathrm{IC}%
\text{\textrm{QC}}\right\rangle =\hat{U}_{\mathcal{P(SA)}}\left\vert
\mathrm{IC}\text{\textrm{QC}}\right\rangle _{0}$ with
\begin{equation}
\hat{U}_{\mathcal{P(SA)}}=\sum_{p=0}^{4^{n}-1}\left\vert p,\mathcal{P}%
\right\rangle \left\langle p,\mathcal{P}\right\vert \hat{U}_{\mathcal{P}}%
\hat{U}_{\mathcal{SA}}(\hat{s}_p,\hat{\textbf{z}}),\label{icqcU}%
\end{equation}
where the pair observables are defined by $\hat{U}_{\mathcal{SA}}(\hat{s}_p,\hat{\textbf{z}})$
and particularly, $\hat{s}_p$ spans a complete operator set for $\mathcal{S}$ and
$\hat{\textbf{z}}$ denotes $n$ Pauli's operators $\hat{z}$ for $\mathcal{A}$.

Is the ICQC defined above a usual quantum computer merely with more
($n+n_{\mathcal{A}}+n_{\mathcal{P}}=4n$) qubits, but without the
information-completeness and trinary structure? The answer is definitely
\textquotedblleft no\textquotedblright\ because of the conceptual difference
between the two quantum computing devices. To see this, we prepare each qubit
of $\mathcal{S}$ in the initial state $\left\vert +,\mathcal{S}\right\rangle
=\frac{1}{\sqrt{2}}(\left\vert 0,\mathcal{S}\right\rangle +\left\vert
1,\mathcal{S}\right\rangle )$ such that $\left\vert \psi,\mathcal{S}%
\right\rangle $ is in a superposition of all $2^{n}$ bit-values with equal
probability amplitude: $\left\vert \psi,\mathcal{S}\right\rangle =\frac
{1}{\sqrt{2^{n}}}\sum_{x=0}^{2^{n}-1}\left\vert x,\mathcal{S}\right\rangle $.
The initial states of $\mathcal{A}$ and $\mathcal{P}$ are likewise prepared:
$\left\vert \phi,\mathcal{A}\right\rangle =\frac{1}{\sqrt{2^{n}}}\sum
_{y=0}^{2^{n}-1}\left\vert y,\mathcal{A}\right\rangle $ and $\left\vert
\chi,\mathcal{P}\right\rangle =\frac{1}{\sqrt{4^{n}}}\sum_{z=0}^{4^{n}%
-1}\left\vert z,\mathcal{P}\right\rangle $. Such a coherent superposition of
conventional quantum computer's initial states is believed to be the very
reason for the speedup of quantum algorithms \cite{QuInf-book,SciAm}.

In the present case of the simplified ICQC, $\hat{U}_{\mathcal{P(SA)}}$ can
encode information-complete programmed measurement operations
upon $\mathcal{A}$ and $\mathcal{S}$. These operations actually determine
allowed quantum algorithms and their outputs (along the $z$-basis) on $n$-qubit state $\left\vert
\psi,\mathcal{S}\right\rangle $, in the terminology of conventional quantum
computing, in which the usual quantum measurement (also along the $z$-basis)
is now replaced by the programmed entanglement
created by $\hat{U}_{\mathcal{SA}}(\hat{s}_p,\hat{\textbf{z}})$. Then we immediately
see that in the ICQC, one has \textit{dual
parallelism}: Parallelism in initial states as usual \textit{and} parallelism
of programmed operations (algorithms and outputs). In other words, \textit{a
single ICQC with }$4n$\textit{ qubits could compute in parallel $2^{n}$ algorithms
of usual quantum computers with }$n$\textit{ qubits}. Due to this particular
dual parallelism enabled by the ICQC, it is reasonable to expect much higher
computational power with the ICQC.

Actually, the ICQC is, by definition, the most powerful computational machine
on qubit systems in the sense of information-completeness; otherwise it is
information-incomplete. Finding algorithms on the ICQC to explicitly
demonstrate the computational power of the ICQC is surely a future interesting
problem. Also, computational complexity and error-tolerance in the ICQC
framework are two important issues. If nature does use the information-completeness
as a guiding principle, it computes the world we currently know (see
\cite{icQFT} for a further discussion);
such a world could be simulated and thus comprehensible by the ICQC (i.e.,
\textquotedblleft qubitization\textquotedblright\ within an
information-complete trinary description) in principle.

\section{Other conceptual applications}

Below we give, only very briefly, a few conceptual applications of the
information-completeness principle and the ICQT, hoping to shed new light on
some long-standing open questions in physics.

An important question is how to understand the occurrence of the classical
world surrounding us, including the second law of thermodynamics and the arrow
of time, in our new framework characterized by the information-completeness
principle and the trinary picture of nature. Currently we cannot present
quantitative analysis of the problem here---Full answer to the question requires to know
the classical limit of the theory of quantizing gravity+matter, which is a highly
nontrivial problem even in loop quantum gravity \cite{Rovelli-book,Thiemann}. Yet,
a qualitative and conceptual answer
to the problem is quite transparent: For information-complete quantum
systems, interactions lead to $\mathcal{P}$-$\mathcal{SA}$ entanglement and
the programmed $\mathcal{S}$-$\mathcal{A}$ entanglement; the Universe as a
whole has an increasing entanglement, a kind of \textit{entanglement arrow of
time} (see also Ref. \cite{icQFT}). It is easy to verify the entanglement
creation by considering the $\mathcal{PSA}$ evolution governed by
$\mathcal{\hat{I}}_{\mathcal{PSA}}$ from a separable state. At a
thermodynamic/macroscopic scale, tracing out thermodynamically/macroscopically
irrelevant degrees of freedom, only as an approximate description of the
underlying information-complete physics, leads to the second law of
thermodynamics, the arrow of time, and ultimately, the classical world.

We note related analysis on the role of entanglement in the thermodynamic
arrow of time in the framework of conventional
\cite{arrow-Maccone,arrow-Terry} or time-neutral formulation \cite{Hartle} of
quantum mechanics. As gravity arguably plays an essential role in our
information-complete description of nature, it is intriguing to see that
gravity plays some role in the occurrence of the second law of thermodynamics
and the arrow of time, as hinted in the study of black-hole thermodynamics
\cite{Bekenstein,BekensteinPRD,Hawking,Thiemann}. In the Di\'{o}si-Penrose
model \cite{Diosi,Penrose}, gravity was argued to play certain role for the
wave function collapse.

Now let us briefly consider the potential conceptual applications to
cosmology. Obviously, the conceptual difficulty of applying usual QT to the
whole Universe disappears in our ICQT. Actually, the ICQT is
a \emph{self-explaining quantum structure} and does not need an observer
as the observer is a part of the Universe. The Universe described
by the ICQT is thus also self-explaining: The constituent parts in trinity are mutually
defining and measuring in a specific dual entanglement structure, eliminating
any subjective aspects regarding the current interpretations of quantum
states---The existence of the Universe does not rely on the existence of
potential observers observing the Universe. Entanglement in the dual form
encodes, without relying on any external observers, all physical information
and can give all physical predictions of the theory.

There is no reason why we cannot describe our human beings as an
information-complete (classical, but ultimately, quantum) system via a
trinary description. In this way, some aspects of human beings could be
comprehensible purely from the informational and physical point of view. For
instance, if we define Alice's body and all of her sense organs as
$\mathcal{A}$ and her outside world as $\mathcal{S}$, then Alice knows her
world or gets known by her world via interaction (i.e., information exchange)
between $\mathcal{S}$ and $\mathcal{A}$. Now an intriguing problem arises
here: What is the programming system $\mathcal{P}$ in this context? A
straightforward way is simply to define $\mathcal{P}$ (or, the correlations
between $\mathcal{P}$ and $\mathcal{SA}$) as the mind (consciousness). By
analogy to the above quantum trinary description, the mind $\mathcal{P}$ and
$\mathcal{SA}$ are mutually defined in an information-complete sense. This
prescription thus provides an interesting possibility of understanding the
most mysterious part (namely, consciousness) of human beings from an
informational and physical perspective. Particularly, Alice's brain undertakes
only \emph{partial} (though the most important) functioning of her mind
according to the above definition; the remaining functioning of her mind is
distributed nonlocally in such a way that enables programming the interaction
between $\mathcal{S}$ and $\mathcal{A}$. Note that, in an
information-complete field theory \cite{icQFT}, the programming system is
the quantized spacetime. If we take the above analogy more seriously, a very
strange conclusion seems to be unavoidable: Alice's mind $\mathcal{P}$ should
be ultimately explainable by physical spacetime $\mathcal{P}$, namely, the mind is
certain (nonlocal) spacetime code of Alice's $\mathcal{SA}$. The reason behind
the conjecture is the strong belief that information-completeness should
underlie the world, ranging from the elementary trinity (elementary fermions,
their gauge fields, and spacetime) to our human beings and the whole
Universe---actually, everything in the world is built from the
information-complete elementary trinity; the matter-spacetime trinity is
an indivisible single entity.

In certain sense, it seems that human beings work as a quantum-decohering
ICQC. Similarly to the fact that an ICQC is conceptually different from a
normal quantum computer, a classical computing device with an integrated
trinary structure similar to the ICQC should be quite different from the
normal Turing machine and could be capable of simulating human-like
intelligence better. Does this mean certain ``consciousness computing'', or
``intelligence computing''? Further consideration in the context will be given
in future.

Therefore, it could well be that information-completeness is of
significance in a broader sense and should be a basic requirement for any
physical systems, classical or quantum. It is in this sense that
information-completeness deserves to be named as a principle. It is a missed
principle in our current understanding of nature and a rule behind the
comprehensibility of the world---The information-complete world is
comprehensible by information-complete human beings.

\section{Conclussions and outlook}

In the present work, we have presented an interpretation-free QT under the
assumption that quantum states of physical systems represent an
information-complete code of any possible information that one might
access. To make the information-completeness explicitly in our formalism,
the trinary picture of describing physical systems seems to be necessary.
Physical systems in trinity evaluate and are entangled both in a dual form;
quantum entanglement plays a central role in the ICQT---Our world is
information given in terms of entanglement at the most fundamental level. So
\textit{the ICQT modifies two postulates (on quantum states and on dynamics)
of current quantum mechanics in a fundamental way and eliminates the
measurement postulate from our description; as a result of the modifications,
the observables can be either information-complete (for }$\mathcal{P}%
$\textit{ or }$\mathcal{SA}$\textit{) or information-incomplete (for
}$\mathcal{S}$\textit{ or }$\mathcal{A}$\textit{)}. We give various evidences
and conceptual applications of the ICQT, to argue that the ICQT, naturally
identifying gravity as nature's programming system in the field-theoretic
case, might be a candidate theory capable of unifying matter and gravity
(spacetime) in an information-complete quantum framework; for further
development on our theory in the context of quantum gravity coupled with
matter, see Ref.~\cite{icQFT}. In this sense, the conventional QT will be an
approximation of our ICQT when quantum effect of gravity is ignored. Such an
approximation leads to the approximate Schr\"{o}dinger equation and the
probability description of current QT. This is in the exact sense that
classical Newtonian mechanics is an approximate description of relativistic
systems. The ICQT motivates an interesting application to
information-complete quantum computing.

As we argued above, current quantum mechanics is \textit{not}
information-complete because of its classical-quantum hybrid feature and
thus, suffers from interpretational difficulties. The explicit demand of
information-completeness not only removes the conceptual problem of our
current understanding of quantum mechanics, but also leads to a profound
constraint on formulating quantum theory. Thus, the ICQT should not be
understood simply as another interpretation of current QT; rather, it, by
giving up the classical concept of probability associated with the measurement
postulate, generalizes current quantum formalism---the physical prediction
(outcomes of an observable and the corresponding probabilities) of a quantum
measurement in conventional QT is now entailed by entanglement; no
entanglement implies no information and thus no prediction. As we noted
previously, adding information-completeness requirement into our current
quantum formalism leads to serious consequences: Information-completeness
not only restricts the way on how to describe physical systems, but also the
way how they interact/entangle with each other \cite{gGUT}. This will thus
give a very strong constraint on what physical processes could have happened
or be allowed to happen.

On one hand, the ICQT provides a coherent conceptual picture of, or sheds new
light on, understanding some problems or phenomena in current physics,
including the intrinsic trinity of matter fermions, gauge fields and gravity,
the occurrence of the classical world, the arrow of time, and the holographic
principle. On the other hand, some other problems, such as the complementarity
principle, quantum nonlocality \cite{Tipler} and quantum communication, should
be reconsidered from the viewpoint of the ICQT. All current quantum
communication protocols \cite{QuInf-book,BB84,teleport-PRL} have to make use
of classical concepts on information. It is thus very interesting to see how
to do communication in the ICQT and, particularly, to see whether or not it is
possible to achieve unconditionally secure communication.

According to the ICQT, the world underlying us is all about information
(entanglement); it is information-complete, deterministic, self-defining,
and thus objective. Such a world view (\textquotedblleft quantum
determinism\textquotedblright) is of course quite different from what we learn
from current quantum mechanics, but in some sense, returns to Einstein's world
view and not surprisingly, represents an embodiment of Wheeler's thesis known
as \textquotedblleft it from bit\textquotedblright\ \cite{Wheeler}. Such a
viewpoint calls for a reconsideration of our current understanding on physical
reality, information, spacetime (gravity), and matter, as well as their links.
Let us cite the famous Einstein-Podolsky-Rosen paper \cite{EPR} here:
\textquotedblleft\textit{While we have thus shown the wave function does not
provide a complete description of the physical reality, we left open the
question of whether or not such a description exists. We believe, however,
that such a theory is possible}.\textquotedblright\ It is too early to judge
whether or not our ICQT completes current quantum mechanics in the
Einstein-Podolsky-Rosen sense cited above, as experiments will be the ultimate
judgement. But if nature does work like a description provided by the ICQT,
nature will be very funny and more importantly, nature does be comprehensible
via a self-defining quantum structure. Einstein might be very happy to see that two of
his important theoretical achievements, namely, general relativity (after
being quantized in modern language) and the concept of quantum entanglement
(discovered by him, together with Podolsky and Rosen), are very essential for
our information-complete quantum description.

\ \ \newline\textbf{Acknowledgements}\newline I am grateful to Chang-Pu Sun
for bringing Refs.~\cite{Tipler,objectQM} into my attention, and to Xian-Hui
Chen, Dong-Lai Feng and Yao Fu for enjoyable discussions. I also acknowledges
University of Science and Technology of China, where the work was initiated.

\end{document}